\documentclass[aps,prb,reprint,superscriptaddress,longbibliography,floatfix]{revtex4-2}

\usepackage{hyperref}
\usepackage{amsbsy,amssymb,amsmath,bm}
\usepackage{graphicx,color,epsfig,rotate}
\usepackage{subfigure}
\usepackage{float}
\usepackage{xspace}
\usepackage{placeins}
\usepackage{ulem}
\usepackage{xcolor}

\newcommand {\UTe}{UTe$_\mathrm{2}$\xspace}
\newcommand {\Tc}{$T_\mathrm{c}$\xspace}

\begin{document}

\title{Slow magnetic quantum oscillations in the \texorpdfstring{$c$}{c}-axis magnetoresistance of \texorpdfstring{\UTe}{UTe2}}

\author{Freya Husstedt}
\affiliation{Hochfeld-Magnetlabor Dresden (HLD-EMFL) and W\"urzburg-Dresden Cluster of Excellence ct.qmat, Helmholtz-Zentrum Dresden-Rossendorf, 01328 Dresden, Germany}
\affiliation{Institut f\"ur Festk\"orper- und Materialphysik, Technische Universit\"at Dresden, 01062 Dresden, Germany}

\author{Motoi Kimata}
\affiliation{Institute for Materials Research, Tohoku University, Sendai, Miyagi, 980-8577, Japan }

\author{Sajal Naduvile Thadathil}
\affiliation{Hochfeld-Magnetlabor Dresden (HLD-EMFL) and W\"urzburg-Dresden Cluster of Excellence ct.qmat, Helmholtz-Zentrum Dresden-Rossendorf, 01328 Dresden, Germany}
\affiliation{Institut f\"ur Festk\"orper- und Materialphysik, Technische Universit\"at Dresden, 01062 Dresden, Germany}

\author{Beat Valentin Schwarze}
\affiliation{Hochfeld-Magnetlabor Dresden (HLD-EMFL) and W\"urzburg-Dresden Cluster of Excellence ct.qmat, Helmholtz-Zentrum Dresden-Rossendorf, 01328 Dresden, Germany}

\author{Markus K\"onig}
\affiliation{Max Planck Institute for Chemical Physics of Solids, 01187 Dresden, Germany}

\author{Gerard Lapertot}
\affiliation{Univ. Grenoble Alpes, CEA, Grenoble-INP, IRIG, PHELIQS, 38000 Grenoble, France}

\author{Jean-Pascal Brison}
\affiliation{Univ. Grenoble Alpes, CEA, Grenoble-INP, IRIG, PHELIQS, 38000 Grenoble, France}

\author{Georg Knebel}
\affiliation{Univ. Grenoble Alpes, CEA, Grenoble-INP, IRIG, PHELIQS, 38000 Grenoble, France}

\author{Dai Aoki}
\affiliation{Institute for Materials Research, Tohoku University, Oarai,
Ibaraki, 311-1313, Japan}

\author{J. Wosnitza}
\affiliation{Hochfeld-Magnetlabor Dresden (HLD-EMFL) and W\"urzburg-Dresden Cluster of Excellence ct.qmat, Helmholtz-Zentrum Dresden-Rossendorf, 01328 Dresden, Germany}

\author{Toni Helm}
\altaffiliation{Corresponding author: \textbf{t.helm@hzdr.de}}
\affiliation{Hochfeld-Magnetlabor Dresden (HLD-EMFL) and W\"urzburg-Dresden Cluster of Excellence ct.qmat, Helmholtz-Zentrum Dresden-Rossendorf, 01328 Dresden, Germany}
\affiliation{Max Planck Institute for Chemical Physics of Solids, 01187 Dresden, Germany}

\date{\today}

\begin{abstract}
Details of the electronic band structure in unconventional superconductors are key to the understanding of their fundamental ground state.
The potential spin-triplet superconductor \UTe, with $T_\mathrm{c}\approx 2.1\,$K, has attracted attention recently.
Its main Fermi surface consists of weakly corrugated, two-dimensional Fermi-surface cylinders that run along the crystallographic $c$ axis. 
In addition, there is evidence for the presence of an additional small three-dimensional band.
This has been discussed controversially as it may be essential for the realization of  superconductivity in \UTe.
Here, we investigate the angle-resolved magnetoresistance and Hall effect in bulk crystalline samples with current along the $c$ axis in fields up to $60\,$T.
We observe low-frequency magnetic quantum oscillations with light effective masses that are most pronounced for magnetic field applied along the $a$ axis.
Two distinct frequencies indicate two separate changes in the Fermi-surface topology, likely connected with Lifshitz transitions.
We discuss the origin of these oscillations in terms of magnetic breakdown, quantum interference, and other potential mechanisms.

\end{abstract}

\maketitle

\section{Introduction}
Key to unraveling the nature of novel quantum phases in unconventional metals is a detailed knowledge about their electronic band structure~\cite{Shoenberg_1984}.
Since the discovery of unconventional superconductivity in \UTe in 2019~\cite{RanScience2019,AokiJPSJ2019}, evidence for a spin-triplet ground state in this material has been reported from various experiments.
A few examples of these are as follows: Enhanced upper critical fields beyond the Pauli-paramagnetic limit were observed~\cite{AokiJPCM2022};
an anisotropic spin susceptibility was probed by nuclear magnetic resonance spectroscopy~\cite{Nakamine2021,Fujibayashi2022OrderParam, Ambika2022,Matsumura2023}; and
high magnetic-field re-entrant superconductivity was detected in pulsed-field experiments surviving up to 70\,T for specific directions~\cite{Ran2019extreme, KnafoCommPhys2021, Lewin2023, Helm2024, Wu2024}.
Yet, the overall ground state of \UTe remains enigmatic and under intense debate~\cite{AokiJPCM2022, Xu2019FermiSurface, IshizukaPRL2019, Shaffer2022, Machida2020, Yarzhemsky2020, Kittaka2020, Shishidou2021, Machida2021,Eo2022, Shick2021, Moriya2022, Choi2024correlated, Hayes2024}, notably regarding the potential of a topological chiral superconducting ground state~\cite{Ran2019extreme, Helm2024, Lewin2023}.
Moreover, novel mechanisms may play a leading role in its field-enhanced superconducting state with distinct order parameters between the low field-low pressure phase and the high field-high pressure phase~\cite{Ran2019extreme, RosuelPRX2023, Helm2024, Vasina2024}.
Magnetic field and its orientation with respect to the crystal lattice play an important role in the behavior of \UTe.
Field-induced changes of the Fermi surface (FS) are indeed expected for different field orientations:
A metamagnetic transition was observed for a field above 35\,T aligned along the $b$ axis~\cite{Miyake2019, MiyakeA2022, NiuPRR2020, KnafoJPSJ2019, RanScience2019};
changes in thermo-electric transport indicated a potential Lifshitz transition at 6\,T for $H \parallel a$~\cite{NiuPRL2020}.

The electronic band structure of \UTe has been intensely investigated by multiple methods such as photoemission spectroscopy~\cite{Fujimori2019, Miao2020, Fujimori2021, Shick2021} and x-ray spectroscopy~\cite{Liu2022, Wilhelm2023, Christovan2024}, providing first insights on the FS and the influence of $5f$ electrons in \UTe.
All experiments point to an intermediate valence state close to 2.5, even though the degree of localization and the exact valence state of the uranium in \UTe are still under debate.
Besides two cylindrical electron- and hole-like quasi-two-dimensional (quasi-2D) FS sections, additional three-dimensional (3D) FS pockets were reported~\cite{Fujimori2019, Miao2020, Broyles2023}. 

Magnetic quantum oscillations (MQOs), so-called de Haas--van Alphen (dHvA) oscillations, detected by torque magnetometry~\cite{Aoki2022dHvA} and Shubnikov--de Haas (SdH) oscillations, detected by proximity-/tunnel-diode-oscillator (P/TDO) measurements~\cite{Eaton2024quasi, Broyles2023} and magnetoresistance (MR)~\cite{Aoki2023}, confirmed the presence of warped 2D FS cylinders.
Moreover, additional low-frequency MQOs were detected by the PDO and TDO method that did not match the large quasi-2D FSs~\cite{Broyles2023, Weinberger2024}.
These low-frequency MQOs might be associated with small FS pockets.
The presence or absence of such a small FS pocket, with heavy or light band masses, is essential for understanding the superconducting order parameter in \UTe~\cite{Choi2024correlated, Sato2017}.
The experimental evidence for such small FSs, however, is being discussed controversially:
While one experiment implied a 3D pocket from the detection of slow MQOs that were observed for any orientation of the external field~\cite{Broyles2023}, others found signatures of such low frequencies only for particular directions~\cite{Weinberger2024}.
In the latter case, mainly due to the corresponding low effective masses, the slow MQOs have been associated with quantum interference between sections of two distinct quasi-2D FS cylinders~\cite{Weinberger2024}. 

Nevertheless, alternative origins for the observed MQOs may still be possible for quasi-2D FSs:
One of the most prominent effects that can lead to the emergence of additional frequencies is magnetic breakdown between trajectories belonging to different electronic bands~\cite{Shoenberg_1984}.
Also, more exotic effects, such as oscillations in the chemical potential~\cite{Steep1999}, quasiparticle-lifetime oscillations~\cite{Huber2023quantum,Leeb2025}, or slow oscillations from weakly warped 2D FSs~\cite{Kartsovnik1988, Kartsovnik2002}, may account for low-effective-mass frequencies similar to what is reported for \UTe.
Moreover, the presence of additional 3D FS pockets may enable more complex scenarios for possible trajectories via the MB mechanism observable in MQO studies.

To date, the low-frequency quantum oscillations have only been observed in the electrical transport of \UTe and not in dHvA measurements ~\cite{Aoki2022dHvA,Aoki2023}. 
Notably, the previous reports on the observation of slow oscillations from small FS sheets relied on results obtained with the P/TDO technique, which mainly were sensitive to the conductivity on the surface of the sample.
It is, therefore, essential to confirm the reported MQOs with a bulk method.

Here, we investigate the angle-resolved magnetoresistance and Hall effect in bulk crystalline samples with current along the $c$ axis in fields up to $60\,$T.
We investigate the temperature and angle dependence of low-frequency MQOs of single-crystalline \UTe in steady and pulsed magnetic fields up to 24 and 60\,T, respectively.
We find two distinct frequencies, $F_1\approx 90\,$T and $F_2\approx 205\,$T, that emerge consecutively.
These oscillations show up for a field aligned within a narrow angular region around the crystallographic $a$ direction.
$F_2$ matches the previously reported frequency values obtained from P/TDO measurements associating the oscillations with quantum interference.
Our results imply as-yet unreported Lifshitz transitions that affect the Fermi-surface topology. 
We discuss the origin of these oscillations in terms of magnetic breakdown (MB), quantum interference, and other potential mechanisms.

\section{Methods}

\textbf{Crystal growth:}
\UTe single crystals were prepared by the molten-salt-flux (MSF) growth technique at Tohoku University, Oarai (Japan) and CEA, Grenoble (France), respectively.
Sample S1 is a bulk sample with \Tc$=2.1\,$K, polished down to a thickness of about $100\,$\textmu m, which was contacted using gold wires and spot welding, exhibiting a residual resistance ratio (RRR) of about 24.
The contacts showed less than $1\,\Omega$ resistances.
Sample S2 is a focused-ion-beam (FIB) microfabricated Hall-bar structure (with a $T_\mathrm{c}$ of 1.9\,K and RRR $\approx 15$) cut from a single crystal with \Tc$=2\,$K, determined by specific-heat measurements.
To prevent degradation due to exposure to air, the structure, including the contacts, was encased in an insulating epoxy.
The encapsulation may introduce additional pressure to the structure that could be the reason for the slightly reduced \Tc.
The cross section of S2 was $(w\times t)=(2.0\times4.0)$\,\textmu m$^2$ with a length of $l=24\,$\textmu m between the voltage contacts.
A second sample, S3, was cut from the same crystal with $I\parallel a$ that exhibited a \Tc of about 2\,K and a RRR of 105 in resistance measurements.

\textbf{Magnetotransport measurements:} 
We measured the resistance using a standard AC four-point lock-in technique.
We performed steady-field measurements on sample S1 in a 24\,T and 15\,T superconducting magnet, both equipped with a two-axis rotator and a $^3$He insert at the Institute for Materials Research of the Tohoku University in Sendai, Japan. 
We also conducted steady-field measurements on sample S2 in an 18\,T superconducting magnet equipped with a single-axis rotator and a top-loading dilution refrigerator, in a 16\,T superconducting magnet equipped with a single-axis Attocube (piezo-active) rotator and a $^3$He insert, and in a pulse magnet, providing fields up to 60\,T, at the Dresden High Magnetic Field Laboratory in Dresden-Rossendorf, Germany.

\textbf{DFT calculations:}
We reproduced previous density functional theory (DFT) calculations~\cite{IshizukaPRL2019} utilizing the full-potential local orbit minimum basis code (\textsc{FPLO}, version 22.00-26)~\cite{Koepernik1999}.
We used the general gradient approximation (GGA)~\cite{Perdew1996} and applied an additional, repulsive Coulomb potential on the 5f electrons of the uranium atoms a few eV.
This potential moved the otherwise itinerant $5f$ electrons into localized states. We conducted all calculations as fully relativistic to take the impact of spin-orbit coupling into account.
We used a $k$-mesh with $20\times 20\times 20$ points to calculate the density of states and band structure and a finer grid with 40 (80 for $U=2\,$eV) points per dimension for calculating the FS.

\section{Results and Discussion}
\begin{figure}[tb]
	\centering
	\includegraphics[width=0.9\columnwidth]{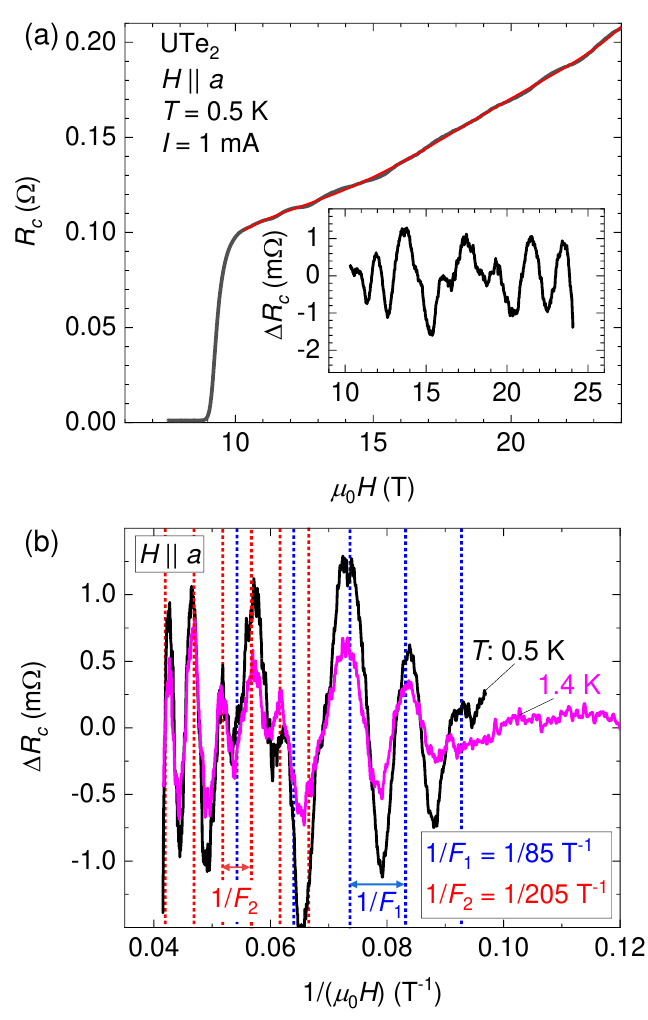}
	\caption{
    (a) $c$-axis MR ($I\parallel c$) of the bulk sample S1 recorded at 0.5\,K for $H\parallel a$ with $I=1\,$mA. Inset: Residual MR after subtraction of a fit with polynomial of fifth degree (red line in main panel) to the data between 10 and 24\,T.
    (b) Residual MR after background subtraction (polynomial of fifth degree) for $T=0.5$ and $1.4\,$K, respectively, plotted against inverse magnetic field $H\parallel a$.
    Blue and red dashed lines indicate the $1/(\mu_0H)$ oscillation periods for the respective frequencies $F_1$ and $F_2$.
    }
	\label{figA}
\end{figure}
We investigated two different samples of a similar high quality, labeled S1 and S2 (for sample details see the experimental section above).
We measured the MR of S1 at various temperatures between 0.6 and $3\,$K, with $I\parallel c$ and the field oriented along $a$.
The overall normal-state MR increased monotonically with increasing field, at least at low temperature, see Fig.~\ref{figA}(a).
Slow SdH oscillations, discernible already with the bare eye, emerged above 10\,T [see also the background-subtracted residual-MR curves in the inset of Fig.~\ref{figA}(a) and in Fig.~\ref{figA}(b)].
Apparently, there are two distinct oscillation periods, $1/F_1$ and $1/F_2$ [red and blue dashed lines in Fig.~\ref{figA}(b)], indicative of more than one oscillation frequency, but emerging consecutively during the field sweep below and above 16\,T, respectively.
In general, MQOs are expected to be periodic in $1/H$.
However, we find that beyond 16\,T, $1/F_1$ does not match with the oscillation maxima/minima any longer and the smaller amplitude oscillations with $1/F_2$ emerge.

\begin{figure}[tb]
	\centering
	\includegraphics[width=1\columnwidth]{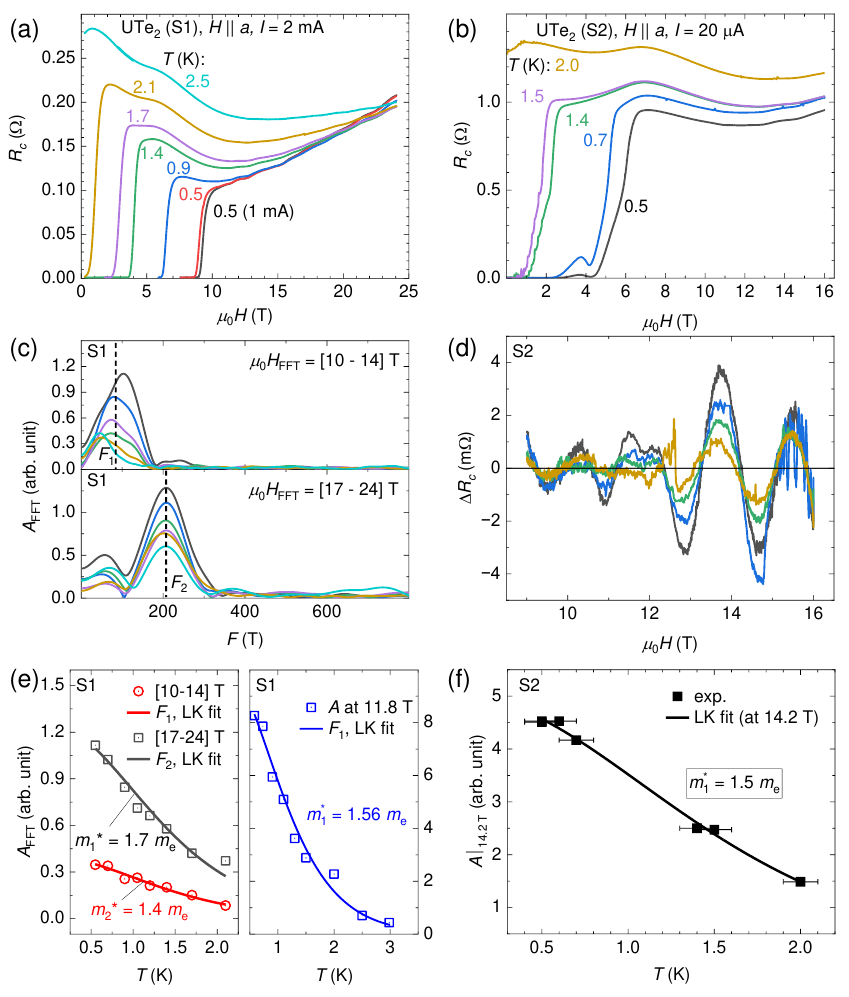}
	\caption{
    (a), (b) $c$-axis resistance of S1 and S2, respectively, recorded in two different setups at various temperatures for field aligned parallel to the $a$ axis.
    (c) FFT spectra of the residual MR of S1 obtained from the data shown in (a) for two different field ranges, 10-14\,T and 17-24\,T, respectively.
    (d) Residual background-subtracted MR (fifth degree polynomial) of S2 from data in (b) plotted against magnetic field.
    (e) Effective-mass plots, Left: FFT amplitudes for S1 versus temperature for $F_1$ and $F_2$ for two field windows 10-14\,T and 17-24\,T, respectively. Right: Amplitude at 11.8\,T for S1.
    (f) Effective-mass plot for the mean amplitude values between the maximum and minimum around 14.2\,T according to Eq.~(\ref{eqLK}) for S2.
    }
	\label{figB}
\end{figure}

In Figs.~\ref{figB}(a) and \ref{figB}(b), we present further MR data recorded for the two samples, S1 and S2, at various temperatures ranging from 0.5 to 2.5\,K.
For both samples, a negative MR with a hump-like feature at around 6-7\,T appears in the normal state upon increasing the temperature.
This hump was already observed in previous magnetotransport studies that connected it to a feature observed simultaneously in the Hall as well as in the Seebeck effects~\cite{NiuPRL2020}.
It was ascribed to a Lifshitz transition, which suggests a significant change in the electronic band structure.  
In the fast-Fourier-transformation (FFT) spectra for the low- and high-field range [Fig.~\ref{figB}(c)], peaks at $F_1\approx85$\,T and at $F_2\approx200$\,T show up.
Note that as the two oscillations exhibit different amplitudes and are best discernible within separate field ranges below or above 16\,T, we chose two distinct field windows, namely, 10-14\,T and 17-24\,T, for the analyses of the temperature damping of $F_1$ and $F_2$, respectively.
Moreover, we describe the temperature dependence of the FFT amplitudes according to the Lifshitz-Kosevich (LK) damping term:
\begin{equation}
R_\mathrm{T}=\frac{\alpha m^*T/B}{\sinh{(\alpha m^*T/B)}},  
\label{eqLK}
\end{equation}
with $\alpha=-14.69\,$T/K and $m^*$ being the effective electron mass.
Both frequencies are damped almost equally by temperature, indicating low effective-mass values of approximately $1.5(2)m_\mathrm{e}$, where $m_\mathrm{e}$ is the free-electron mass.
The fits of the mass plots are presented in Fig.~\ref{figB}(e), left panel.
We, furthermore, conducted measurements in a 15\,T magnet system and confirmed the effective mass value for S1 with $m_1^*\approx1.5(1)m_\mathrm{e}$ for $F_1$ [see Fig.~\ref{figB}(e), right panel].

A similar MQO to $F_1$ is observable in the transport signal up to 16\,T of the FIB microfabricated sample S2 [Fig.~\ref{figB}(d)].
The determined effective-mass value is $1.5(1)m_\mathrm{e}$ [cf. mass plot in Fig.~\ref{figB}(f)], so basically the same as that obtained for S1. 
Even though the microstructured sample exhibits lower critical fields for $H\parallel a$, the oscillations look very similar with approximately the same low frequency of about $F_1=90\,$T.
Interestingly, for S2, we observe an additional higher-order variation in the residual MR, visible by two local maxima in the oscillation near 12\,T [Fig.~\ref{figB}(d)].
The split maximum seems not to be present in the data for S1.
Assuming a periodicity according to the frequency $F_2$, the split maximum at 12\,T cannot be associated with that frequency.
As we already described above for sample S1, the periodicity of $F_1$ seems to change in the vicinity of 16\,T, just before it is suppressed completely.
Hence, the variation in the periodicity of $F_1$ for sample S2 may have a similar origin.

\begin{figure}[tb]
	\centering
    \includegraphics[width=0.8\columnwidth]{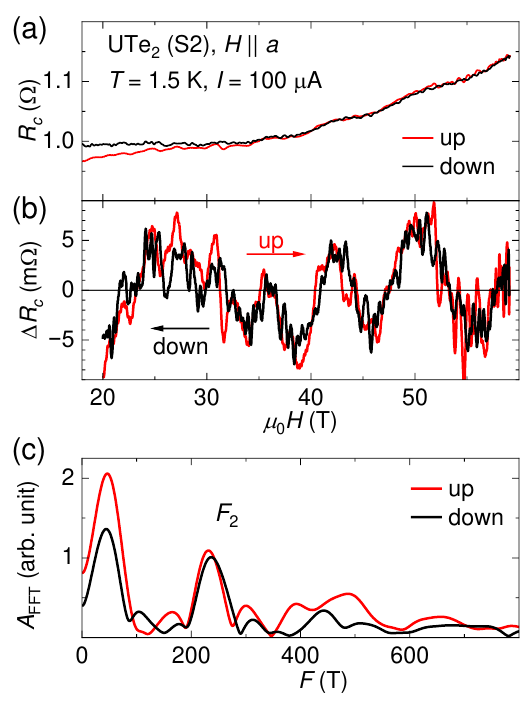}
    \caption{
    (a) MR of S2 recorded at 1.5\,K in a pulsed magnet up to 60\,T with $H\parallel a$.
    A slight heating, likely caused by eddy currents, is responsible for the deviation between the up and down sweep.
    (b) Residual MR after background subtraction of a cubic polynomial. 
    (c) FFT spectra for a field window from 20 to 59\,T.
    Note: The peak at 50\,T strongly depends on the choice of the polynomial degree and is therefore not related to MQOs. 
    }
	\label{figC}
\end{figure}
In an attempt to trace the oscillations towards higher fields, we study S2 down to 1.5\,K in a pulsed-field setup up to 60\,T, see Fig.~\ref{figC}.
Owing to the enhanced contact resistances, the sample receives a significant heat load from the application of a current of $100\,$\textmu A required for a reasonable signal-to-noise ratio.
The comparison to the steady-field data indicates a sample temperature of about 2.5\,K.
In the pulsed-field MR data in Fig.~\ref{figC}(a), we see a deviation between the up and down sweeps at low fields that originates from additional heating during the pulse, potentially related to eddy currents.
Moreover, the high-resistance contacts forced us to apply a relatively low excitation frequency of 11\,kHz.
With an approximate pulse rise time of 30\,ms and a pulse decline within 100\,ms, this further limited the resolution during the experiment.
Nevertheless, we were able to resolve MQOs, as can be seen from Fig.~\ref{figC}(b).
For $H\parallel a$, the FFT spectrum exhibits a clear peak at around $F=210(5)\,$T, which agrees within error bars with the value of $F_2$ determined for S1 in steady fields up to 24\,T.

\begin{figure*}[tb]
	\centering
	\includegraphics[width=0.95\textwidth]{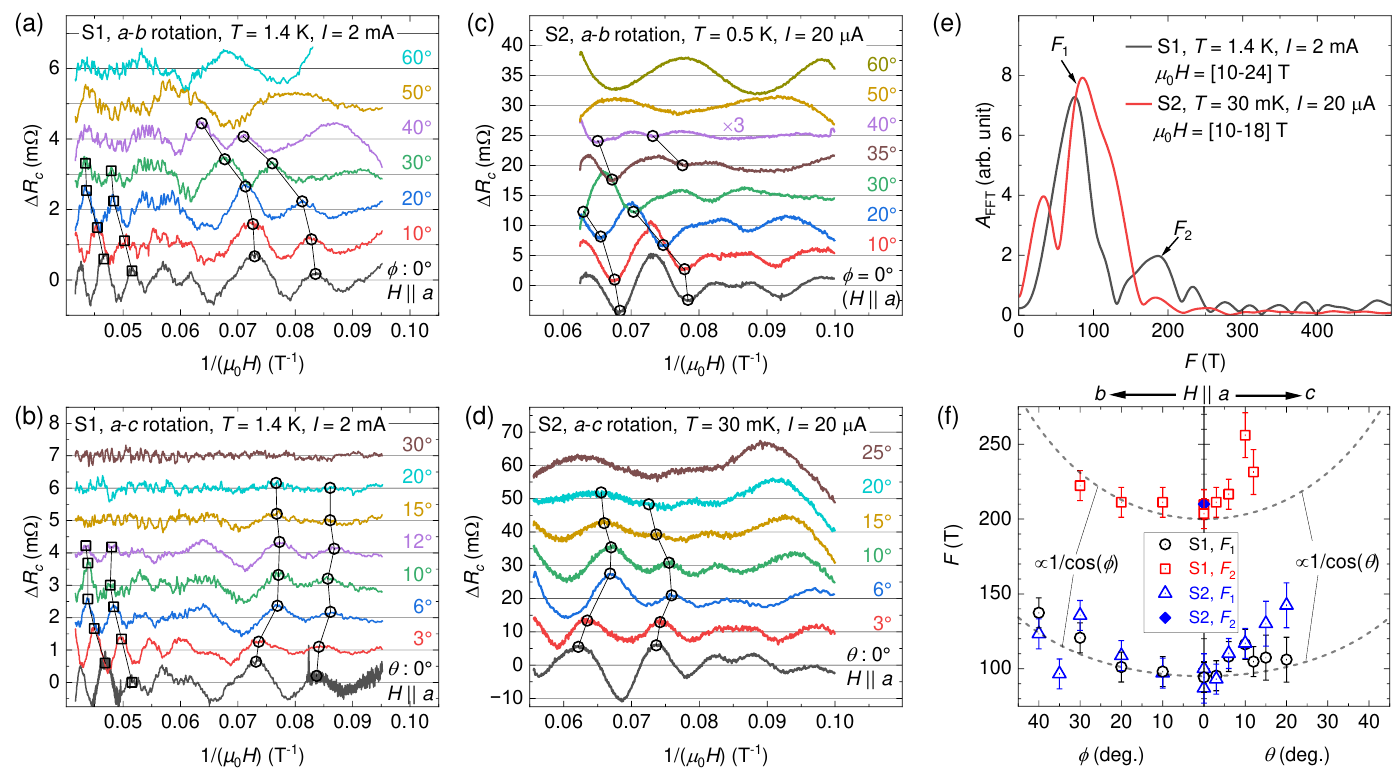}
	\caption{
    (a),(b) Residual MR (background subtraction using a polynomial of fifth degree) recorded for S1 at $1.4\,$K with $I=2\,$mA  for various tilt angles within the $a$-$b$ and $a$-$c$ rotation plane, respectively.
    Note: Curves in (a)--(d) are shifted by constant offsets for better visibility.
    (c),(d) Residual MR (background subtraction using a polynomial of fifth degree) for S2 measured at various tilt angles at 0.5 and 0.03\,K, respectively, with $I=0.02\,$mA, recorded during two separate experiments in a 16\,T and an 18\,T magnet system.
    (e) FFT spectra for S1 and S2 and $H\parallel a$.
    (f) Frequency values determined from the periodicity of the oscillations as indicated by circles and squares in (a)--(d) for sample S1 (black circles and red squares) and S2 (blue triangles).
    The blue diamond marks the data point obtained in 60\,T for $H\parallel a$. 
    }
	\label{figD}
\end{figure*}
In the following, we will turn to the angle dependence of the observed MQOs.
We measured MR of S2 up to 60\,T for $H\parallel b$, $H\parallel c$, and $40^\circ$ tilted away from $b$ towards $c$. 
We could not resolve any MQOs, at least within the noise level that was similar to the one in Fig.~\ref{figC}.
We were able to trace oscillations within the $a$-$b$ (azimuthal angle $\phi$) and the $a$-$c$ (polar angle $\theta$) plane for both samples in steady fields in separate setups and laboratories.
In  Figs.~\ref{figD}(a) to \ref{figD}(d), we present selected residual MR curves obtained at various tilt angles.
To trace the frequencies $F_1$ and $F_2$ for different angles, we extract the periodicity between the maxima and minima (marked by black circles and squares) and plot them for both samples in Fig.~\ref{figD}(f).
As can be seen in Fig.~\ref{figD}(e), an analysis based on a FFT yields broad peaks due to the low number of maxima and minima within the given field ranges.
This, in particular, leads to a broadening of the FFT peaks in the low-frequency range between 0 and approximately 50\,T, which is close to the value of $F_1$.
Moreover, the amplitudes of the peaks in the FFT spectra are very sensitive to the background subtraction and the degree of the chosen polynomial.
We also checked lower-degrees polynomials for subtracting the background and made sure that no frequencies were removed or added.
Here, we determine the periodicity $1/F$ between the observed maxima, as indicated by the black circles in Figs.~\ref{figD}(a) to \ref{figD}(d)

Let us first focus on the bulk sample S1 [Figs.~\ref{figD}(a) and \ref{figD}(b)].
Both $F_1$ and $F_2$ are clearly traceable for azimuthal tilt angles up to $\phi\approx 40^\circ$ from $a$ towards the $b$ axis.
By contrast, for polar orientations within the $a$-$c$ plane, the oscillations are already suppressed at $\theta>20^\circ$.
We also show selected data for S2 in Figs.~\ref{figD}(c) and \ref{figD}(d). 
For sample S2, there are clear slow oscillations with a frequency of approximately 90\,T (matching $F_1$ observed in sample S1) that also get suppressed for $\phi>40^\circ$.
The frequency increases slightly when rotating towards $a$ and towards $c$.
The oscillations vanish beyond $\theta\approx 20^\circ$.
Both frequencies do not strictly follow $1/\cos\theta$ (dashed lines), which is the dependence expected for a 2D FS.

As mentioned above, for sample S2 there is an additional variation in the oscillation maximum at about 12\,T.
This feature is responsible for a broadening and splitting of the FFT peak in Fig.~\ref{figD}(e).
Apparently, for this sample, the periodicity in the observed MQOs labeled with $F_1$ changes slightly between 10 and 16\,T, indicative of a more complex behavior.
However, the small number of discernible maxima and minima limits more detailed analyses.

While a frequency of about 90\,T has not been reported to date, $F_2$ matches the 200\,T frequency previously observed in the surface conduction for $H\parallel a$~\cite{Eaton2024quasi, Weinberger2024}.
There, these MQOs were associated with Stark oscillations that were caused by quantum interference between different sections of the FS~\cite{Stark1977}.
\begin{figure}[tb]
	\centering
	\includegraphics[width=0.8\columnwidth]{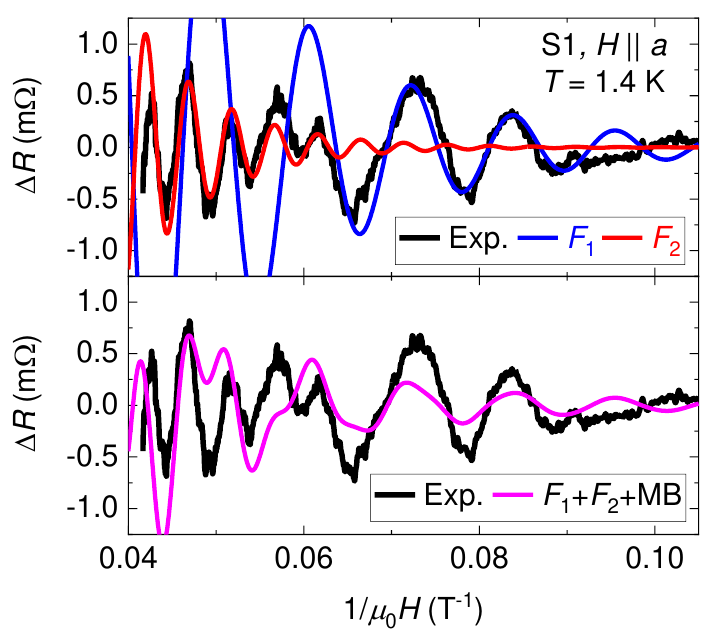}
	\caption{
    (Black curve in both panels) Same data as shown in Fig.~\ref{figA}(b).
    (Blue curve) Simulation according to the LK formula, Eq.(\ref{eqLK}), with $F=85\,$T  and $m^*=1.5\,m_\mathrm{e}$.
    (Red curve) Simulation according to the LK formula with $F=205\,$T and $m^*=1.5\,m_\mathrm{e}$.
    (Magenta curve) Superposition of both frequencies with parameters from both simulations above, plus MB damping factors $R_{1,\mathrm{MB}}$ and $R_{2,\mathrm{MB}}$, with $B_\mathrm{MB}=6\,$T.
    }
	\label{figE}
\end{figure}

Besides quantum interference, the MB effect is a candidate that could explain the sudden onset of $F_2$ and the fading out of $F_1$~\cite{Kaganov1983}.
In the MB scenario, charge carriers can overcome a small gap, $\Delta_\mathrm{MB}$, between adjacent parts of the FS via tunneling for magnetic fields larger than the breakdown field, $B_\mathrm{MB}$.
The field-induced MB trajectories force the charge carriers along different parts of the FS as compared to low fields, which should affect the overall MR as well as the Hall effect depending on the details of the FS~\cite{Falicov1964, Helm2015}.
The gap can be described via the Blount criterion $\hbar\omega_0\sim \Delta_\mathrm{MB}^2/\epsilon_\mathrm{F}$, with $\omega_0=eB_\mathrm{MB}/m^*$ and the Fermi energy $\epsilon_\mathrm{F}$~\cite{Shoenberg_1984}.
In the simplest case, we can consider the LK formula for fixed temperature and field orientation~\cite{Shoenberg_1984}:
\begin{equation}
\Delta \rho = B^{1/2}\sum_{i=1,2} A_i R_{i,\mathrm{D}} R_{i,\mathrm{MB}}R_\mathrm{i,S}\cos{\left[2\pi \left(\frac{F_i}{B}+\gamma_i\right)\right]},
\end{equation}
where $A_i$ are oscillation amplitudes for each frequency, $R_{i,\mathrm{D}}= \exp{(\alpha m_i^*T_\mathrm{D}/B)}$ are Dingle damping factors, $R_{i,\mathrm{MB}}$ are MB damping factors, $R_\mathrm{i,S}$ are spin damping factors, and $\gamma_i$ oscillation phases.
In Fig.~\ref{figE}, we demonstrate this by a simple simulation using a superposition of the two frequencies, $F_1=85\,$T and $F_2=205\,$T, shown in the upper panel as blue and red curves.
We introduce the damping factors $R_{1,\mathrm{MB}}=\left[1-\exp{\left(-B/B_\mathrm{MB}\right)}\right]^{n/2}$ and $R_{2,\mathrm{MB}}=\exp{\left(-B/B_\mathrm{MB}\right)}^{n/2}$ with the number of breakdown junctions $n$, which we assume to be 2 (see below).
A small breakdown field of $B_\mathrm{MB}=6\,$T is sufficient to yield a curve that is roughly similar to the experimental one (magenta curve in Fig.~\ref{figE} lower panel).
Although the MB effect may account for the emergence of $F_2$, it cannot explain the sudden suppression of $F_1$ as observed. 
Our results indicate that $F_1$ vanishes very abruptly around 16\,T and remains completely suppressed for higher fields.
In a conventional MB scenario, the two frequencies are expected to transition smoothly as explained above.

\begin{figure}[tb]
	\centering
	\includegraphics[width=1\columnwidth]{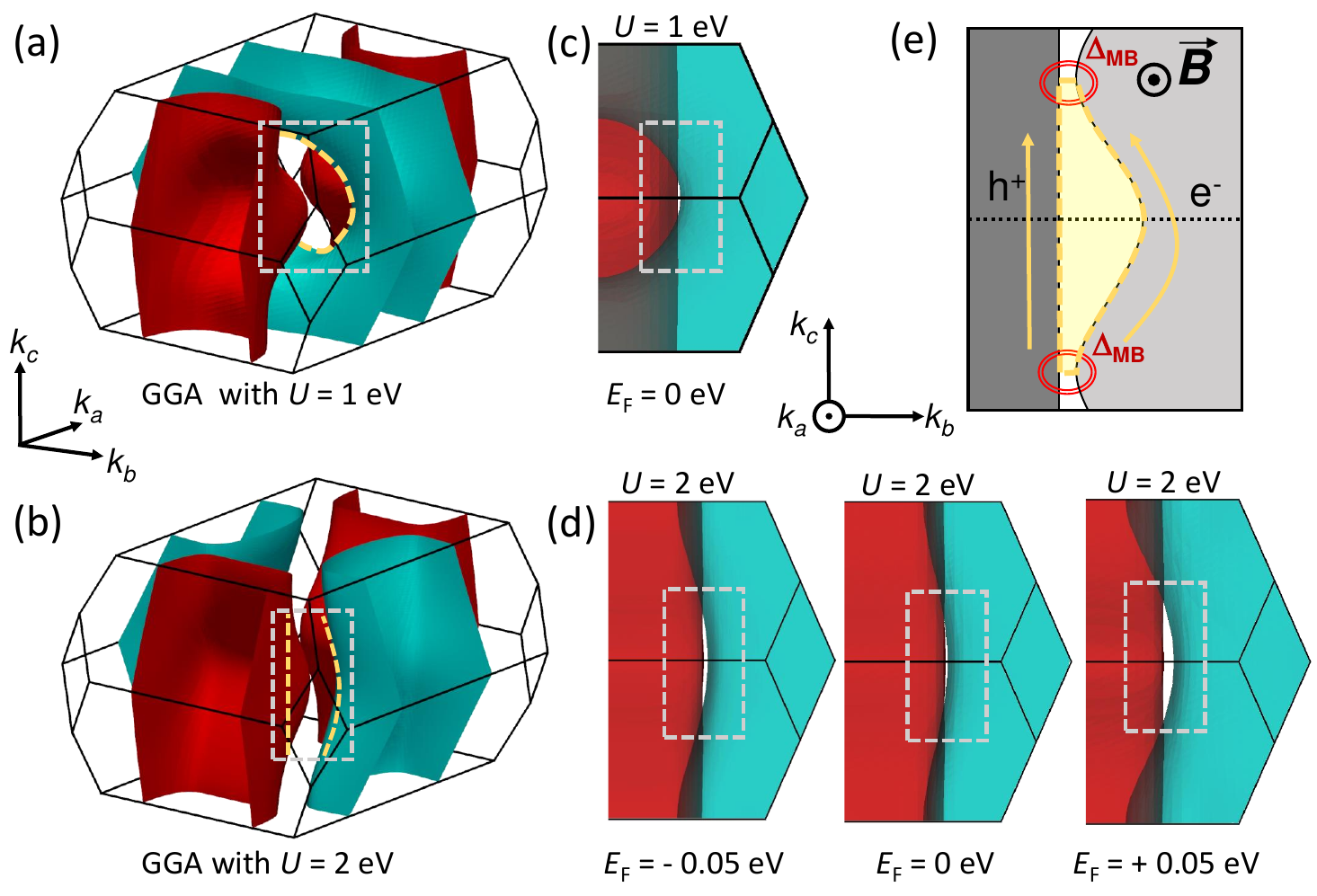}
	\caption{
    (a),(b) DFT calculated FS of \UTe with (blue) electron- and (red) hole-like FS sheets for the $U$ values 1 and 2\,eV, respectively.
    (c),(d) 2D projection of the FSs in (a) and (b) along the $k_a$ direction.
    (d) 2D projection for three different values of the chemical potential, 0 and $\pm0.05\,$eV.
    (e) Schematic of quantum interference between two charge-carrier trajectories (marked by yellow arrows) along the $k_c$ direction separated by a gap, $\Delta_\mathrm{MB}$, which is narrow enough to enable MB.
    }
	\label{figF}
\end{figure}

Let us now look at the FS of \UTe.
In Figs.~\ref{figF}(a) and \ref{figF}(b), we show results from DFT calculations for two different $U$ values of 1 and 2\,eV.
We chose the latter value to reproduce previously reported results~\cite{IshizukaPRL2019} and the former to showcase the development of a 3D pocket around the center of the Brillouin zone.
For smaller $U$, the electron band exhibits a more 3D character with a closed loop around the $\Gamma$ point in the first Brillouin zone [marked by the yellow dashed line in Fig.~\ref{figF}(a)].
Hence, such a closed trajectory could account for a MQO observable for $H\parallel a$.
Nevertheless, the result shown in Fig.~\ref{figF}(b), exhibiting two quasi-2D FS cylinders, is considered more realistic, since it is in line with recent dHvA measurements~\cite{Aoki2022dHvA,Eaton2024quasi}.
For the predicted quasi-2D FSs oriented along the $c$ direction, an external field applied along $a$ would force the charge carriers along open trajectories [following the vertical yellow dashed lines in Fig.~\ref{figF}(b)], rendering closed cyclotron orbits impossible.
As these bands possess effective masses between 32 and 57 free-electron masses~\cite{Aoki2022dHvA}, we would expect a similar mass value for a conventional MB orbit. The mass should even be further enhanced as the field is tilted towards the in-plane direction, which is clearly not the case neither for $F_1$ nor for $F_2$.
Moreover, if contributions from both electron- and hole-like FSs are involved, the small frequency values may only be possible via classically forbidden trajectories, as the Lorentz force would switch its sign as soon as the charge carriers moved from one to the other band via a potential breakdown gap.
Hence, the observed MQOs for $H\parallel a$ are rather unexpected and likely of an unconventional origin.

Various mechanisms may account for the observed low frequencies: a weak warping of quasi-2D FSs~\cite{Kartsovnik1988, Kartsovnik2002};  oscillations of the chemical potential ~\cite{Steep1999}; inter- and intraband scattering (i.e., quasi-particle lifetime oscillations)~\cite{Leeb2025}; or quantum interference, according to a proposal for \UTe by Eaton \textit{et al.}~\cite{Eaton2024quasi}.

For weakly warped quasi-2D FSs, slow oscillations can emerge that possess extremely low effective masses, as it was reported for organic superconductors~\cite{Kartsovnik1988, Kartsovnik2002}. This effect, however, strongly relies on field orientations aligned along the quasi-2D FS cylinders, which is counter to the observations in this work (strongest MQOs for $I\parallel c$, but with $H\parallel a$).

Oscillations in the chemical potential can be ruled out in our case, as they usually occur with large oscillation amplitudes due to strongly 2D FSs.
Also, these MQOs should be picked up by the magnetization, which, so far, has not been the case here.
Our measurements of magnetic torque on a single crystal (same growth batch of sample S2) and also those of another group~\cite{Aoki2022dHvA}, to date, have only revealed oscillations associated with the quasi-2D FS cylinders.
The slow oscillations appear to be undetectable by these thermodynamic probes, rendering oscillations in the chemical potential an unlikely origin for $F_1$ and $F_2$.
By contrast, oscillations that originate from ntra- or interband scattering or from quantum interference should be most pronounced in electrical-transport probes~\cite{Leeb2025}.

In the first case, quasiparticles may scatter on defects or collective excitations, connecting different parts of the FS.
The corresponding additional MQOs have effective masses that result from a combination of the underlying bands~\cite{Huber2023quantum}.
In the second case of quantum interference, MQOs arise from coherent charge-carrier pathways between different MB junctions, effectively enclosing flux in an interferometer-like configuration~\cite{Shiba1969, Stark1977, Leeb2025}.
MB is a well-known effect in the magnetic field based on tunnel transitions of conduction electrons of a metal between trajectories belonging to different electronic bands~\cite{Kaganov1983, Shoenberg_1984}.
According to the authors of Ref.~\cite{Kaganov1983}, the observation of quantum interference requires a FS that (i)~realizes an interferometer setup, i.e., at least two semiclassical paths on which particles travel from one MB junction to the next and which enclose an area in momentum space; (ii) MB junctions between the semiclassical paths with intermediate MB probabilities $p$ leading to a partial MB with $p(1-p)\neq 0$.
One of the key signatures of quantum interference oscillations (QIOs) is their stability against increased temperature~\cite{Shoenberg_1984}.

According to theory, the effective mass of QIOs results from the difference of the two involved parts of the FS.
Hence, it can be significantly smaller than compared to the fundamental renormalized band masses, reducing the damping of the oscillation amplitude due to temperature.
Specifically, this feature may enable the detection of the MQOs at higher temperatures in the case of the heavy fermion \UTe~\cite{Eaton2024quasi}.

Recent transport studies with current along the $a$ direction confirmed the heavy FSs, but did not resolve these slow MQOs either, at least for a field applied in the $b$-$c$ plane~\cite{Aoki2024}.
We, therefore, tested an additional sample S3 cut from the same crystal used for S2, however, with the current aligned along $a$.
No low-frequency MQOs were discernible for $H\parallel a$ under the same noise conditions at $T=0.6\,$K and field up to 16\,T.
This implies that it is the $c$-axis transport that is most sensitive to the low-frequency MQOs.

Distinguishing quantum interference from intra-/interband scattering is challenging~\cite{Leeb2025}.
In Figs.~\ref{figF}(c) and \ref{figF}(d), we show the 2D projection of the two distinct FS cylinders along the $k_a$ direction.
A small area is discernible where the projections of the two bands do not overlap completely, providing a narrow region where two pathways could lead to MQOs based on quantum interference via the narrow gaps in between~\cite{Weinberger2024}.
QIOs are associated with MB via a narrow gap, $\Delta_\mathrm{MB}$, between two different parts of the FS [Fig.~\ref{figF}(e)] and require a finite tunneling probability.
At large-enough magnetic field, the oscillation amplitude is expected to subside again as the breakdown probability reaches $100\,\%$. Hence, only one path will be allowed for in the interference pattern sketched in Fig.~\ref{figF}(e).

In contrast, the so-called quasi-particle lifetime oscillations (QPLOs) originate from intra- and interband scattering processes and would exhibit ever-growing amplitudes with increasing field~\cite{Leeb2023}.
In our high-field data, we do not observe a significant damping of the oscillation strength up to 60\,T, that is, for $F_2$ the oscillation amplitude increases monotonically upon increasing field up to 60\,T [Fig.~\ref{figC}(b)].
Moreover, QPLOs would be very stable against temperature and can enable closed orbits between distant parts of a FS, rendering them a likely alternative origin for the observed MQOs.
An additional point against the suggested quantum interference is that the orbits, required for MQOs when $H\parallel a$, run along the heavy open quasi-2D bands and are connected via MB gaps only at specific touching points.
Therefore, very few charge carriers would be expected to contribute to the $c$-axis magnetotransport, which would result in an overall negligible contribution to the main conduction channel.
However, at this point, making a clear distinction between QIOs and QPLOs goes beyond the scope of this work.

In Fig.~\ref{figF}(d), we present DFT results with an artificial variation in the position of the chemical potential by $\pm50\,$meV.
This demonstrates that the overlap between the projections of the two bands strongly depends on the details of the electronic structure, which certainly are beyond the reach of the present state of DFT calculations in \UTe.
Note that other calculations based on dynamic mean-field theory show significant differences to the DFT results~\cite{Choi2024correlated, Halloran2024}.
As we observe an abrupt transition from $F_1$ to $F_2$ induced by field, the question arises how magnetic field may affect the FS properties and what mechanism can account for the observed transition. 
Even though QIOs or QPLOs seem plausible, at this point, we cannot rule out a more complex FS topology that may even include a smaller 3D FS pocket responsible for the low-frequency MQOs $F_1$ and $F_2$.

\begin{figure}[tb]
	\centering
	\includegraphics[width=0.8\columnwidth]{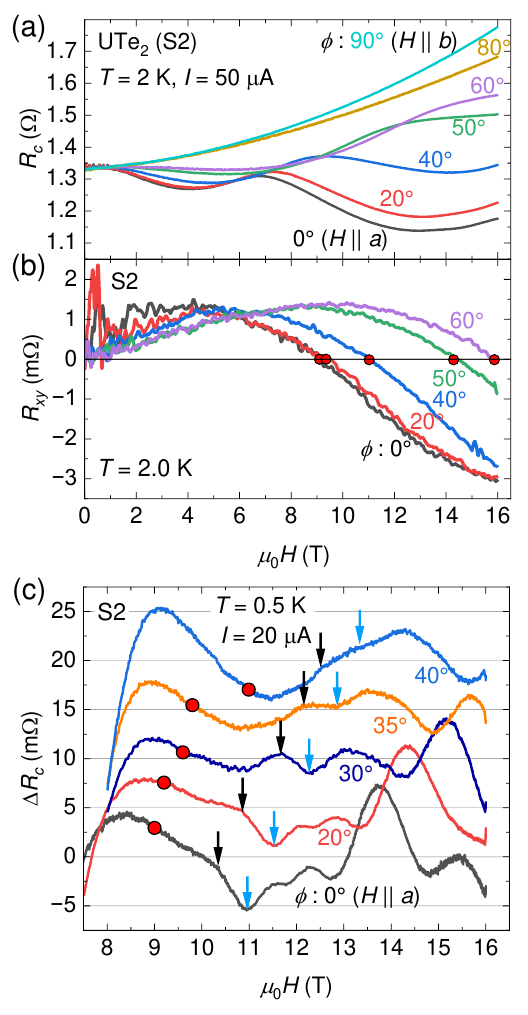}
	\caption{
    (a) $c$-axis MR recorded for various fixed angles $\phi$ with 20\,\textmu A excitation current at $2\,$K.
    (b) Hall resistance recorded simultaneously with the data shown in (a).
    Red dots mark zero crossings. 
    (c) Residual MR for $T=0.5\,$K after subtracting the slowly varying background fitted by a polynomial of fourth degree.
    Red dots on the flat part before the oscillations emerge mark the values where the Hall resistance, shown in (b), changes its sign.
    The initial upturn originates from the superconducting transition at low fields. 
    Black arrows highlight the first hump-like features.
    Blue arrows highlight the first minima.
    }
	\label{figG}
\end{figure}

The presence of a 3D FS is supported by the small anisotropy of the electrical transport in \UTe~\cite{Eo2022, Thebault2022}.
Moreover, the overall MR for $H\parallel a$ appears to be relatively small, e.g., at 0.5\,K it changes by a factor of 2 between 10 and 24\,T [Fig.~\ref{figA}(a)].
This is surprisingly weak for a quasi-2D FS with open bands along the $c$ axis.
Such open bands should result in a non-saturating MR for $H\parallel a$ and a strongly anisotropic MR between this orientation and $H\parallel c$.
Even though for $H\parallel a$ the MR exhibits no sign of saturation, we observe an equal or even smaller MR than for the other two principle axes.

$F_2$ is very similar to the  MQO previously reported by Weinberger \textit{et al.}~\cite{Weinberger2024}.
Even the angular dependence agrees rather well.
For rotations from $a$ towards $c$, we resolve oscillations up to about $20^\circ$ tilt, matching almost perfectly with the previous report.
In the case of rotation from $a$ towards $b$, we are able to trace the oscillation up to $40^\circ$.
In Ref.~\cite{Weinberger2024}, additional data for rotation within the $a$-$b$ plane is provided.
There, a faint peak is still resolvable in the FFT spectra for tilts larger than $20^\circ$ up to $28^\circ$.
Even though this is a lower angle as compared to our result, it also reveals a slight anisotropic behavior in the angle dependence of the $\sim200\,$T frequency oscillation.
Hence, the limited angular range for $F_2$ disagrees with the finding in Ref.~\cite{Broyles2023}, where a $\sim 200\,$T frequency was discernible over the full angular range within both the $a$-$b$ and the $a$-$c$ plane, which was associated with a 3D FS pocket.
The apparent anisotropic angle dependence in our case may signal a significant anisotropy in the FS properties.
Our findings, therefore, support the proposed quantum-interference scenario, with interfering trajectories stretched along the $c$ direction [shown in Fig.~\ref{figF}(e)].

In Figs.~\ref{figG}(a) to \ref{figG}(c), we compare MR data obtained for sample S2 for various field orientations within the $a$-$b$ plane at 2\,K with simultaneously measured Hall-effect data and the resolved MQOs superimposed to the MR at 0.5\,K.
There is a broad hump in the normal-state MR for $H\parallel a$ around 6-7\,T [Fig.~\ref{figG}(a)] that shifts towards higher fields upon rotating away from $a$.
The field value of this feature is much lower than the one at which the MQOs oscillations set in [marked by black and blue arrows in Fig.~\ref{figG}(c)].
The Hall resistance exhibits a maximum  and clear sign change (marked by red dots) that both shift to higher fields upon increasing tilt angles $\phi$.
In general, the Hall effect can be comprised of orbital, anomalous, and topological contributions depending on the material~\cite{Nagaosa2010AHE}.
A sign change in the Hall resistivity may have various origins, such as changes in the electronic or magnetic structure.
Previous thermoelectric measurements with $I\parallel a$ and $H\parallel a$ observed features around 6\,T in various properties such as the Seebeck coefficient, the Hall effect, and the MR~\cite{NiuPRL2020}.
A transitional change in the FS properties, associated with a Lifshitz transition at this field value, is proposed.
In general, such a transition is connected with a topological change of the FS, i.e., a change in the charge-carrier distribution at the Fermi edge~\cite{Lifshitz1960anomalies, Yamaji2006}.
Here, the question arises of what effect such a transition has on the FS in \UTe.
The slow oscillations only start emerging at about 11\,T, just above the field range where the Hall effect changes its sign from positive to negative [marked by red circles in Figs.~\ref{figG}(b) and \ref{figG}(c)].
This implies that the FS undergoes significant changes upon increasing field.
Interestingly, Seebeck measurements reveal multiple features at fields of about 6, 10, and 20\,T~\cite{NiuPRL2020}.
Similarly, we observe $F_1$ and $F_2$ to emerge consecutively at different fields in our experiments.

Our results would, therefore, support the presence of at least two field-induced Lifshitz transitions associated with the emergence of $F_1$ and $F_2$.
As we observe differences in the MQOs between the microstructured and bulk sample, pressure induced by the substrate of the epoxy encapsulation may affect the FS of \UTe.
For example, the superconducting ground state in strongly correlated systems may be controllably affected by built-in pressure using microstructures~\cite{Bachmann2019}.
A pressure-dependent study demonstrated pressure-induced changes in the slow MQOs associated with quantum interference~\cite{Weinberger2024pressure}.
Furthermore, a theoretical study highlighted the expected effect of coherent and incoherent Kondo scattering on the FS, causing temperature-, field-, and pressure-induced topological FS transformations~\cite{kang2025coex3d2dfs}.
This proposal could also account for the observed MQOs that emerge and vanish at specific magnetic field values or orientations.

\section{Conclusion}

In summary, we were able to detect slow magnetic quantum oscillations in the resistance of bulk and microfabricated \UTe samples in different experimental setups with samples from different growths.
Two distinct slow frequencies, $F_1\approx 90\,$T and $F_2\approx205\,$T, for a field along the crystallographic $a$ direction, were discernible, that emerged consecutively upon increasing the external magnetic field.
$F_1$ started emerging at around 11\,T, the field range at which the Hall effect crosses zero and changes its sign.
At around 16\,T, $F_1$ ceased to exist and instead a new frequency $F_2$ sets in.

We determined the effective masses of $1.5m_\mathrm{e}$ for both frequencies and traced them depending on tilt angles in the $a$-$c$ and $a$-$b$ plane.
The oscillations were discernible within only a narrow angular region around the $a$ axis.
Although $F_2$ matched previous reports from contactless surface-conductivity measurements~\cite{Weinberger2024}, the lower frequency $F_1$ has not been observed to date.
We argue that these oscillations likely originated from non-Onsager  mechanisms such as quantum interference or quasiparticle lifetime oscillations.

Our observations support the presence of at least two field-induced Lifshitz transitions that alter the FS topology of \UTe. 
Apparently, \UTe undergoes significant changes induced by external magnetic fields both along the $a$ and $b$ axes.
This highlights again the importance of the field as well as its orientation for the complex magnetic phase diagram of \UTe.  
To understand its implications on the superconducting ground state, further evidence for a field-induced Fermi-surface reconstruction and its exact details would be highly desirable.

\section{Acknowledgments}
We thank A. Eaton and M. Kartsovnik for fruitful discussions.
We are grateful for the support by A. Mackenzie and the department of Physics of Quantum Materials at the MPI for Chemical Physics of Solids in Dresden, Germany.
We acknowledge support by the Institute for Materials Research at Tohoku University, Japan.
We acknowledge support from the French National Agency for Research ANR within the projects FRESCO (Project No. ANR-20-CE30-0020) and SCATE (Project No. ANR-22-CE30-0040).
We acknowledge support from the Deutsche Forschungsgemeinschaft (DFG) Grant No. HE 8556/3-1 and the W\"urzburg-Dresden Cluster of Excellence on Complexity and Topology in Quantum Matter---\textit{ct.qmat} (EXC2147, Project No. 390858490).
We acknowledge the support of the HLD at HZDR, member of the European Magnetic Field Laboratory (EMFL).

\bibliography{bibliography}

\end{document}